Giorgio Querzoli[1], Stefania Fortini[2], Stefania Espa[2], Martina Costantini[2], Francesca Sorgini[2]

# Fluid dynamics of aortic root dilation in Marfan syndrome

1. DICAAR, Università degli studi di Cagliari - Via Marengo 3, 09123, Cagliari, Italy

2. DICEA, Sapienza Università di Roma - Via Eudossiana 18, 00184, Roma, Italy

e-mail: querzoli@unica.it

Abstract

Aortic root dilation and propensity to dissection are typical manifestations of the Marfan Syndrome (MS), a genetic defect leading to the degeneration of the elastic fibres. Dilation affects the structure of the flow and, in turn, altered flow may play a role in vessel dilation, generation of aneurysms, and dissection. The aim of the present work is the investigation in-vitro of the fluid dynamic modifications occurring as a consequence of the morphological changes typically induced in the aortic root by MS. A mock-loop reproducing the left ventricle outflow tract and the aortic root was used to measure time resolved velocity maps on a longitudinal symmetry plane of the aortic root. Two dilated model aortas, designed to resemble morphological characteristics typically observed in MS patients, have been compared to a reference, healthy geometry. The aortic model was designed to quantitatively reproduce the change of aortic distensibility caused by MS. Results demonstrate that vorticity released from the valve leaflets, and possibly accumulating in the root, plays a fundamental role in redirecting the systolic jet issued from the aortic valve. The altered systolic flow also determines a different residual flow during the diastole.

## Introduction

Marfan syndrome (MS) is an heritable connective tissue disorder with major manifestations in the ocular, skeletal and cardiovascular systems, mostly related to mutations in the gene coding for fibrillin which causes degeneration of elastic fibres (Pyeritz, 2000). Cardiovascular involvement includes aortic root dilatation and dissection, aortic valve regurgitation, mitral valve prolapse and regurgitation. In fact, mutant fibrillin causes the degeneration of elastic fibres, which results in vessel dilation, decreased thickness and increased stiffness of the arterial walls (Hirata et al., 1991; Savolainen et al., 1992; Teixido-Tura et al., 2014). These factors together concur to the dramatic rise of mural stress: associated complications, such as dissection, are the primary causes of premature death and morbidity (Murdoch et al., 1972; Westaby, 1999). Life expectancy has significantly improved in the last 30 years, mainly because constant monitoring has allowed for effective preventive aortic root replacement (Mimoun et al., 2011). Recently, a central role has been given to aortic root dilation in the diagnostic criteria and, at present, the indication to surgery is mainly based on the measure of the root diameter, which is well known to gradually increase in MS patients (Loeys et al., 2010; Nollen et al., 2004). However, continuous improvement of the criteria for the indication for surgical intervention is crucial for a successful prevention of the complications without exposing patients to unnecessary risks.

On one hand, results of Magnetic Resonance Imaging highlight the altered blood flow characteristics in MS (Geiger et al., 2012). On the other hand, a number of studies indicate that fluid dynamics influences the progression of aortic dilation and aneurysms by changing the spatio-temporal variability of the wall shear stresses which, in turn, could affect endothelial function, and thus promote vascular remodelling (Cecchi et al., 2011; Pries et al., 2005; Reijer et al., 2010), thus closing a feedback loop between fluid dynamics alterations and aortic dilation. Recently, Hope et al. (2013) highlighted the correlation between altered flow and the occurrence of an aortic dissection.

Previous studies (Januzzi et al., 2004; Meijboom et al., 2005) have shown that, compared to other patients with an aortic dissection, dilation occurs more frequently in the aortic root and sinuses of Valsalva in patients with MS. Conversely, dilation in the aortic arch and descending aorta is comparable to that of other patients. Accordingly, dissection of type A (*i.e.* localised only in the ascending aorta, AAo) is more frequent in MS patients.

Aortic flow has been extensively investigated by means of computational fluid-dynamics (De Tullio et al., 2009; Marom et al., 2013; Sturla et al., 2013; Tullio et al., 2011; Yun et al., 2014), which nowadays permits the investigation of patient specific flows, though uncertainty remains in the definition of the constitutive equations of living tissues. Other studies have been performed *in-vitro* granting repeatability of the conditions at the cost of a schematic reproduction of morphology (Brücker et al., 2002; Chandran et al., 1985; Gülan et al., 2012; Hutchison et al., 2011; Leo et al., 2012; Lim et al., 2001; Yoganathan et al., 1979). Recently, investigations have been carried out using passively driven, *ex-vivo* models (Leopaldi et al., 2012; Richards et al., 2009; Vismara et al., 2014), which are optimal in reproducing realistic anatomy but suffers from individual variability and do not permit the application of image velocimetry techniques since they do not allow optical access to the flow. The main purpose of most of the studies about the fluid dynamics of the aortic root was to assess the performances of prosthetic valves. Furthermore, to the extent of the authors' knowledge, no investigations matching the variability of the aortic diameter during the cardiac cycle have been performed so far. Therefore, many aspects of the fluid dynamics of the aortic root in MS remain to be clarified.

The aim of the present study is to elucidate the alteration in the flow structure induced by the aortic root dilation and increased wall stiffness in Marfan patients since they could be potential prognostic markers for the evaluation of the evolution of the disease. Three model aortas made of silicone rubber, representing two typical aortic root dilations and a reference case, have been investigated *in vitro* and results in terms of velocity and vorticity are presented. In addition to geometry, the models reproduced the diameter variation during the cardiac cycle as observed *in-vivo* both in the MS and healthy case. Differences in the flow patterns are shown and discussed.

## 1. Materials and Methods

Figure 1 shows a sketch of the Pulse Duplicator (PD) used during the experiments. It consists of a hydraulic loop where the impedance of vascular systemic net is mimicked following a lumped parameters approach, and the flow is generated by a deformable ventricle (figure 1, V) whose volume is controlled by a piston following an assigned motion law. The PD was initially developed for the study of the intraventricular flow and it is described in detail elsewhere (Domenichini et al., 2007; Espa et al., 2012; Fortini et al., 2013; Querzoli et al., 2010). For the present study, it was equipped with a reproduction of the anatomic district between the left ventricle outflow tract and the aortic root. The model aortic root, made of silicon rubber, was placed inside a parallelepidedal chamber (figure 1, $A_C$ and figure 2) with Plexiglas walls (2.0 cm thick) and filled with water. A bio-prosthetic valve (27 mm, St. Jude Medical Biocor), was housed at the annulus (figure 1, $A_V$). Two piezoelectric transducers probed pressure just upstream and 6.5 D downstream of the valve, namely the ventricular ($p_v$) and aortic ($p_a$) pressure, respectively (D = 27mm indicates the diameter of the AAo). An electromagnetic flow-meter (F) recorded flow-rate just upstream from the aortic valve.

Experimental parameters were tuned to fulfil dynamic similarity with a real world condition of 75 beats per minute and a 64 ml stroke volume (SV) by matching both Reynolds and Womersley number:

$$\text{Re} = \frac{UD}{\nu}; \text{Wo} = \sqrt{\frac{D^2}{T\nu}}$$

where U was the peak aortic velocity, $\nu$ the kinematic viscosity and T the period of the cardiac cycle. Since the geometric scale ratio was 1:1 and the working fluid was water, whose kinematic viscosity ($\nu = 10^{-6}$ m/s$^2$) is about one third that of blood, similarity was obtained by expanding the period three times (T = 2.4 s) while keeping the SV constant at 64 ml.

|    | an [mm] | ra [mm] | aa [mm] | hs [mm] |
|----|---------|---------|---------|---------|
| H  | 25      | 36      | 27      | 20      |
| M1 | 25      | 45      | 27      | 30      |
| M2 | 25      | 48      | 27      | 45      |

**Table 1** Main dimensions of the three model aortic roots. an = aortic annulus, ra = aortic root diameter, aa = ascending aorta diameter, hs = height of sinuses of Valsalva

With these settings, peak velocity was U = 0.20 m/s, therefore, non-dimensional parameters resulted Re = 5400 and Wo = 17. Three silicon rubber aortic root models were moulded and tested (figure 2): the first, resembling the geometry of the healthy aorta (H), the second corresponding to a dilation of the sinuses of Valsalva (M1), and the third with an elongated shape of the dilated sinuses to obtain the typical pear-shaped aorta (M2). Main dimensions of the models are reported in table 1.

In MS degeneration of elastic fibres causes a decreased distensibility of the AAo. This effect was reproduced by the PD, which was designed to control the percentage diameter changes during the cardiac cycle:

$$\Delta\%_{max} = \frac{D_{sys} - D_{dia}}{D_{dia}} \cdot 100,$$

where $D_{dia}$ and $D_{sys}$ are the minimum and maximum diameter of the proximal aorta. The chamber containing the aortic root was connected by means of the resistance Ra (figure 1) to a vertical tube in communication with the atmosphere. As the water of the chamber flows out through the tube, the aortic root dilates, and vice-versa. Therefore, Ra was adjusted to match typical values of $\Delta\%_{max}$ identified by the analysis of *in-vivo* measurements reported in literature (Baumgartner et al., 2006, 2005; Jeremy et al., 1994; Vitarelli et al., 2006): $\Delta\%_{max} = 13$ for the healthy aorta and $\Delta\%_{max} = 6$ for the MS cases. It is worth noticing that, although most *in-vitro* studies described in literature made use of deformable models, no evidence of a quantitative reproduction of the aortic diameter variation during the cardiac cycle was found.

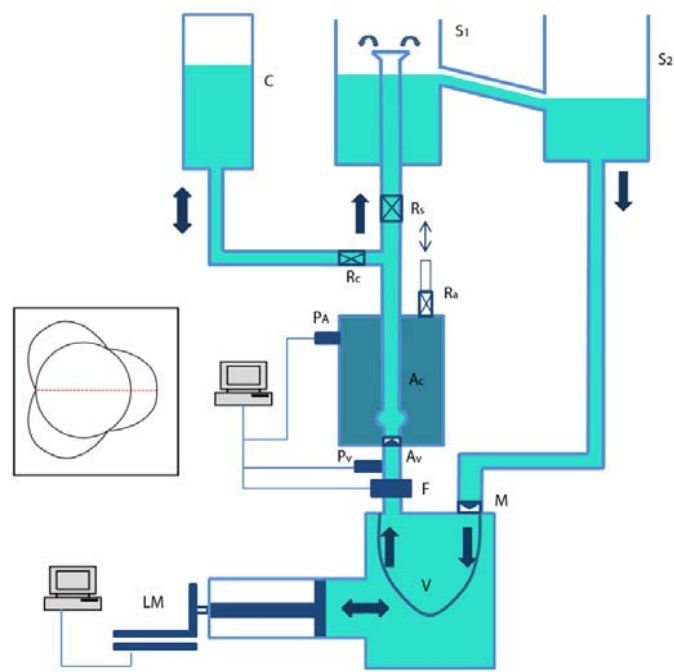

**Fig. 1** Sketch of the Pulse Duplicator. Flow is generated by the alternative motion of the piston driven by the linear motor LM which, in turn, controls the volume of the ventricle V. The ventricle is fed through a check valve M by the constant head reservoir $S_2$, and is connected to the aortic model in the chamber, $A_C$, (dimensions: 0.18m×0.18m×0.3m) made of 0.02m thick Plexiglas walls. $A_C$ is connected to the open air through resistance $R_a$. Flow from the aortic model reaches the reservoir $S_1$. Resistances $R_s$ and $R_c$, together with the air vessel C, mimic the impedance of the systemic circulation. $A_v$: aortic valve; $P_v$ and $P_a$: piezoelectric pressure transducers; F: electromagnetic flow-meter. In the left panel, a short axis section of aortic root. The dashed line indicates the measuring plane.

All measurements where performed during 100 consecutive cycles, and phase averaged data will be presented hereafter. Investigations were carried out on a longitudinal plane of symmetry intersecting the commissure on one side, and the sinus of Valsalva on the other (see the panel on the left of figure 1).

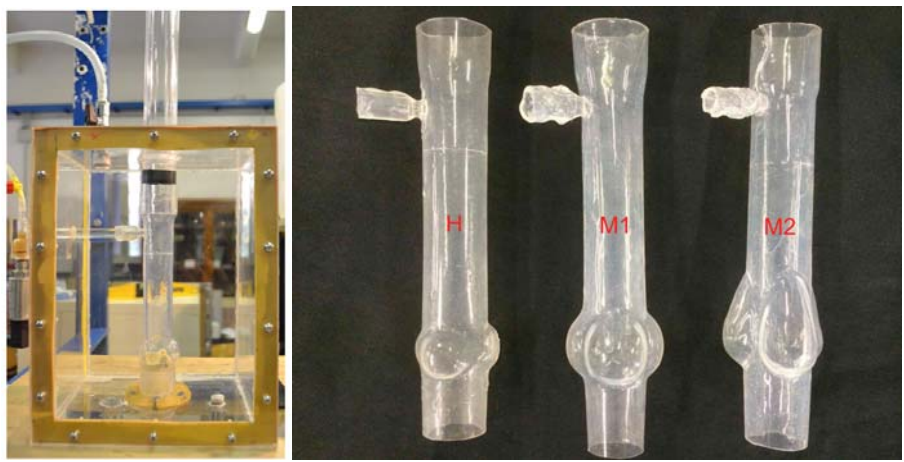

**Fig. 2** On the left: a picture of the aortic chamber. On the right: the model aortas used during the experiments; healthy (H), M1 and M2 from left to right.

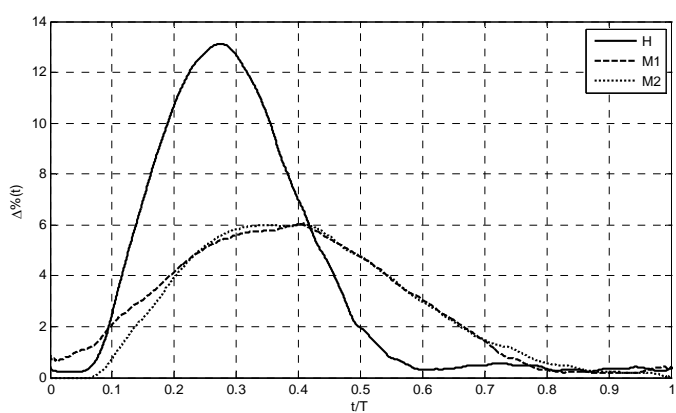

**Fig 3** Percentage diameter change of the proximal ascending aorta during the cardiac cycle $\Delta\%(t) = (D_a(t) - D_{dia}) / D_{dia} \times 100$. Healthy aorta (H): solid line; MS models M1 and M2 are indicated by a dashed and dotted line, respectively.

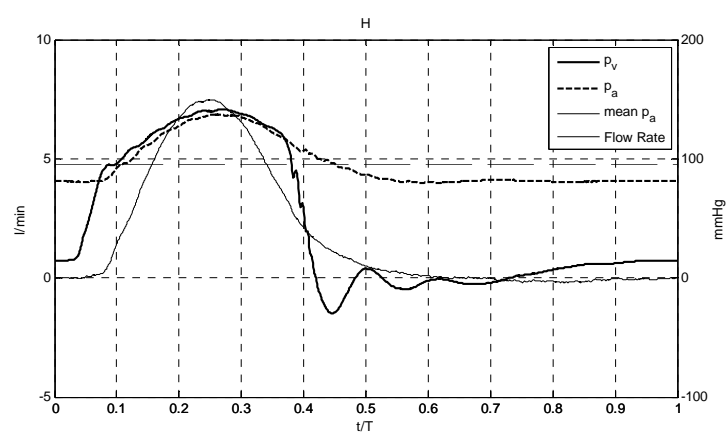

**Fig 4** Aortic ($p_a$) and Ventricular ($p_v$) pressures (scale on the right axis) and flow rate (scale on the left axis) as a function of non-dimensional time in the healthy model (H). A straight dashed line indicates the mean aortic pressure.

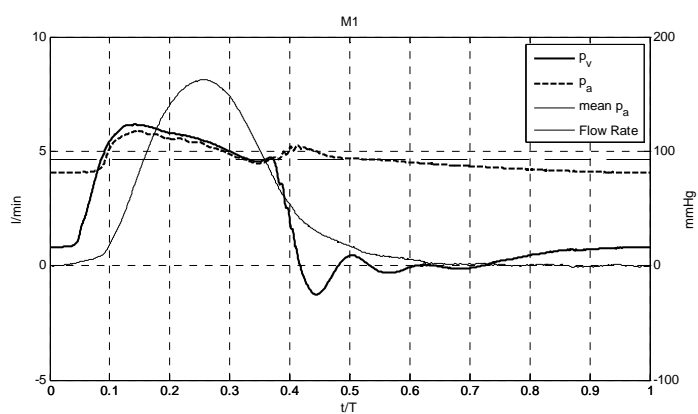

**Fig 5** Aortic ($p_a$) and Ventricular ($p_v$) pressures (scale on the right axis) and flow rate (scale on the left axis) as a function of non-dimensional time in the dilated model M1. A straight dashed line indicates the mean aortic pressure.

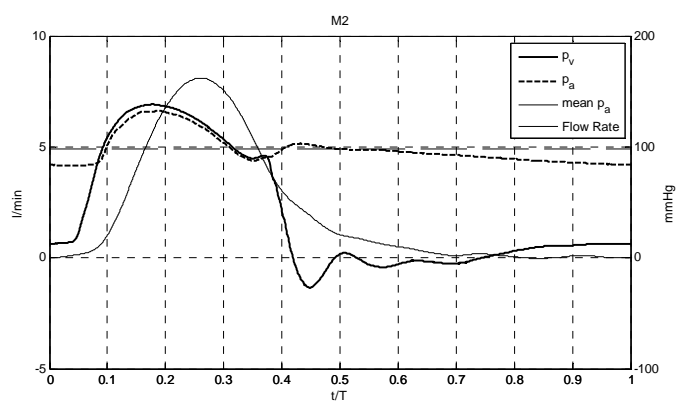

**Fig 6** Aortic ($p_a$) and Ventricular ($p_v$) pressures (scale on the right axis) and flow rate (scale on the left axis) as a function of non-dimensional time in the dilated model (M2). A straight dashed line indicates the mean aortic pressure.

The measuring plane was illuminated by a 12 W, infrared laser and the working fluid was seeded with neutrally-buoyant particles (50 μm average diameter). Images were acquired by a high-speed camera (790 frames per second, 800×950 pixels, 0.06 mm per pixel, resolution). For the computation of the instantaneous aortic diameter, $D_a(t)$, images were thresholded, the walls were recognised, and their distance was computed in the section 2D downstream from the valve.

In figure 3, the measured values of relative diameter variation, Δ%(t), during the whole cardiac cycle are plotted, while figures 4-6 show the ventricular and aortic pressure and flow-rate for the three models. The systemic circulation impedance was tuned to attain in all cases a mean $p_a$ as high as 100 mmHg. Δ%(t) is a maximum at t/T = 0.27, just after the peak flow-rate, in the healthy aorta, whereas the maximum is reached in the range t/T = 0.35 − 0.40 in the models M1 and M2, *i.e.* with a significant delay compared to the peak flow-rate. Feature-Tracking algorithm (Cenedese et al., 2005) was used to measure the instantaneous two-dimensional velocity field by recognizing particle trajectories and interpolating velocity samples on a regular, 50×51, grid. Velocity and vorticity maps obtained by phase averaging over 100 cycles are presented and discussed below.

## 2. Results

In figures 7-9, the velocity fields are shown at three salient points of the cardiac cycle: t/T = 0.17, corresponding to the phase of accelerated systolic ejection, t/T = 0.22 when the flow-rate is a maximum and t/T = 0.50 at the end of the decelerated ejection. The velocity field, described in terms of streamlines (white lines) and velocity magnitude (colour map), is superimposed on an image of the investigation region taken at the same instant and showing the vessel walls (grey). The right wall corresponds to the centre of a sinus of Valsalva, the left corresponds to the commissure (see also the left panel in figure 1). Valve leaflets are clearly apparent on below.

In the healthy model (H), at the accelerated ejection (t/T = 0.17) a jet is entering the aortic root. On the side of the commissure, the shear layer at the border of the jet generates a series of counter-clockwise rotating vortices which are advected downstream. Two of them are visible on the left of the jet in figure 7a. On the opposite side, a clockwise vortex is completely housed by the sinus of Valsalva. The series of vortices shed on the left side is interposed between the jet and the wall throughout the systolic ejection, thus highest velocities are observed on the side of the sinus. As a result, the jet impinges the wall on the side of the sinus just downstream from the sinotubular junction. This scenario is confirmed by the flow field at t/T = 0.22 (figure 7b).

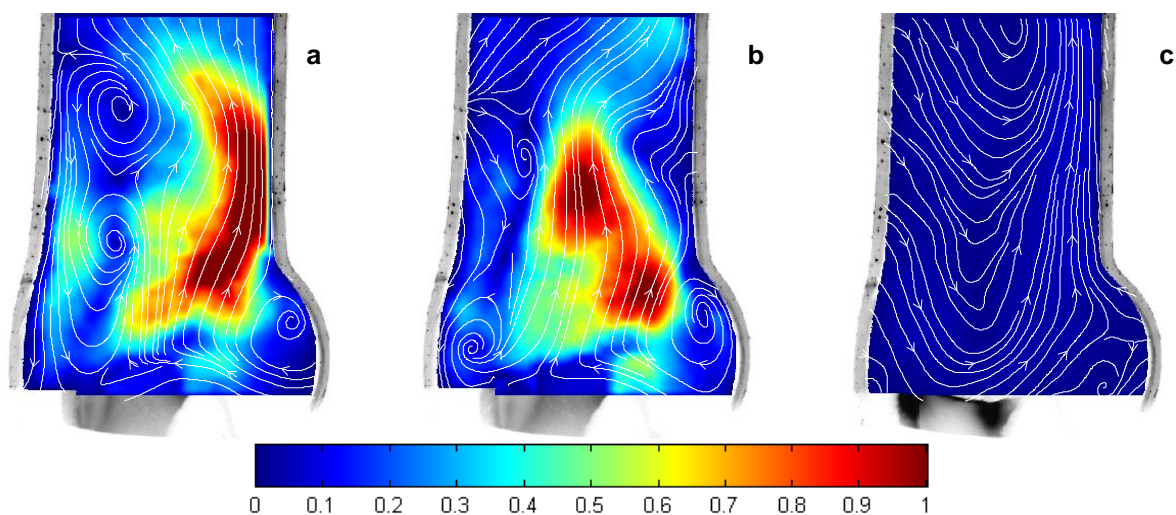

**Fig 7** Instantaneous streamlines in the healthy model, H, at t/T = 0.17 (panel **a**); t/T=0.22 (panel **b**); t/T = 0.50 (panel **c**). Colour indicates velocity magnitude made non-dimensional by the peak velocity U.

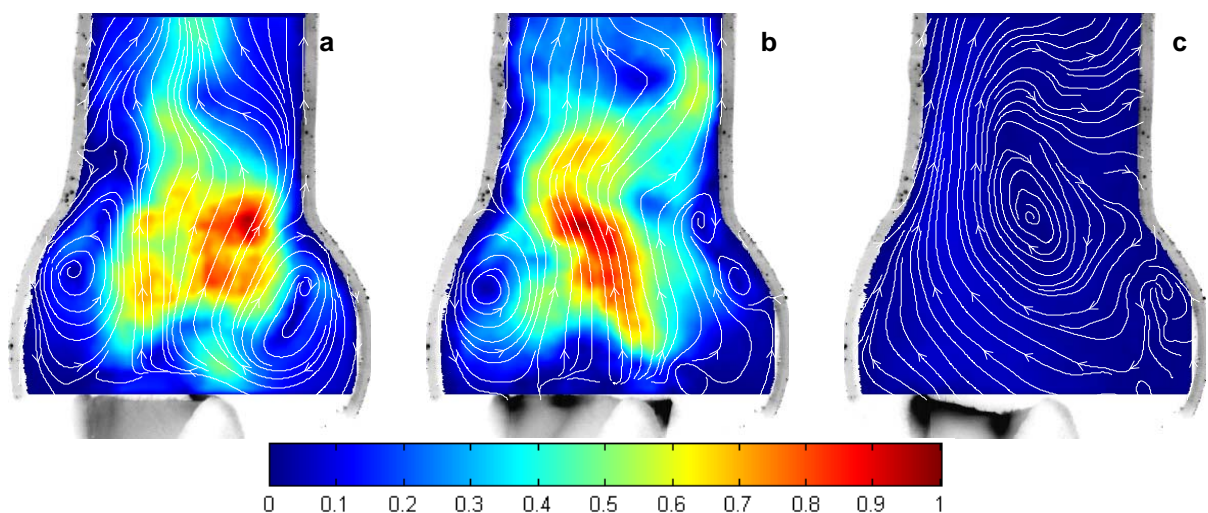

**Fig 8** Instantaneous streamlines field in the model M1 at t/T = 0.17 (panel **a**); t/T=0.22 (panel **b**); t/T = 0.50 (panel **c**). Colour indicates velocity magnitude made non-dimensional by the peak velocity U.

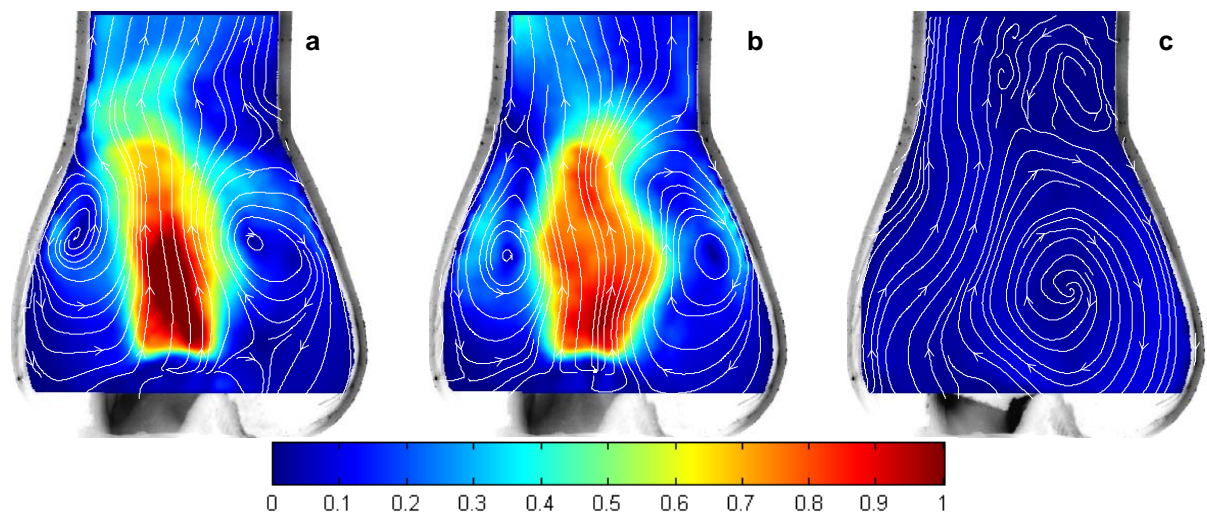

**Fig 9** Instantaneous streamlines in the dilated model M2, at t/T = 0.17 (panel **a**); t/T=0.22 (panel **b**); t/T = 0.50 (panel **c**). Colour indicates velocity magnitude made non-dimensional by the peak velocity U.

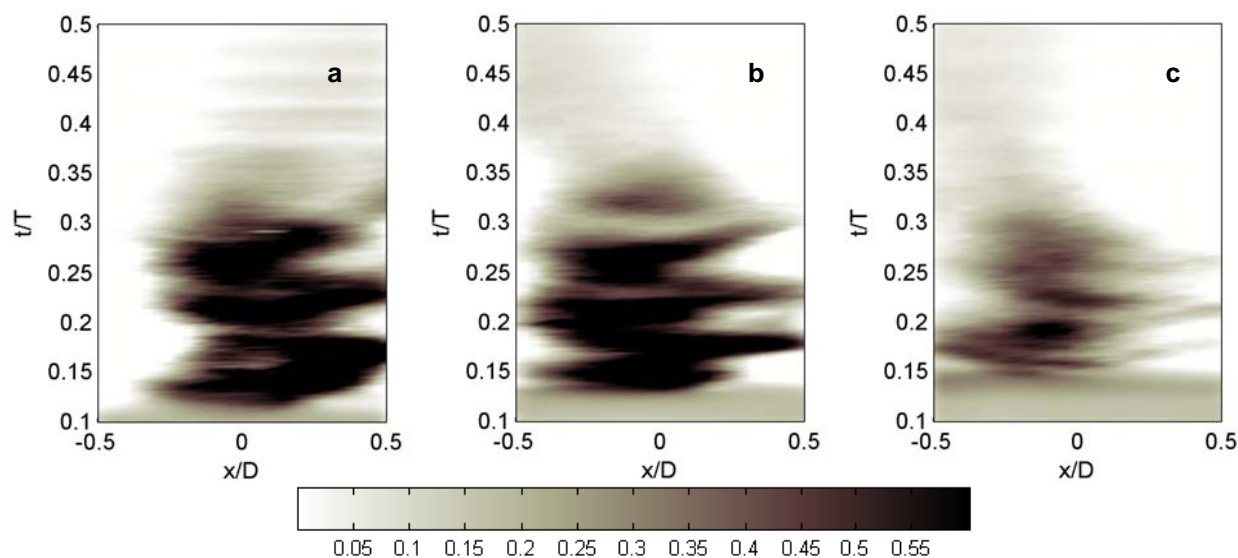

**Fig 10** Velocity magnitude non-dimensionalised by U, on the spanwise section at the level of the sinotubular junction as a function of time. Healthy aorta (H) panel **a**; dilated aorta M1 panel **b**; dilated aorta M2 panel **c**.

Although the main jet is nearly central, a couple of vortices advected downstream are apparent on the left side. At the end of the systole (t/T = 0.5, figure 7c), the residual circulation is mainly counter-clockwise, with the core of the vortex in the AAo, at the downstream end of the investigated region.
In the model aorta M1, a second recirculating vortex develops and remains on the left of the valve lumen both at the accelerated (t/T = 0.17, figure 8a) and decelerated ejection (t/T = 0.22, figure 8b). The jet is bent on the sinus side at the former instant, but it impinges the left wall at the latter instant. That last orientation of the jet prevails and generates a general clockwise circulation, centred at the height of the sinotubular junction, at the end of the systole (t/T = 0.5, figure 8c).
Also in the model aorta M2, a second vortex resides on the commissure side during the whole systolic ejection (figure 9a, and 9b). No vortex shedding is observed in this case. In this model the sinus is more elongated, thus a larger recirculation region finds room on the right side. Its effect seems to prevail during the whole systolic ejection thus the jet orientation is leftwards both at t/T = 0.17 and t/T = 0.22. At the end of the systole (figure 9c), the clockwise circulation of the right vortex prevails and dominates the whole aortic root.

Compared to the residual circulation characterising model M1, the centre of rotation is more proximal.
The time evolution of the oscillation of the jet is described in figure 10. Colour maps give a spatiotemporal representation of the velocity magnitude on the cross-flow section at the level of the sinotubular junction during the systole. The jet oscillates throughout the accelerated ejection in all the configurations. This seems to be the two-dimensional signature of the precession of the jet which possibly occurs in pipe flows through a sudden expansion (Cafiero et al., 2014). Later, its orientation remains stable on one side. The period of the oscillations is about 0.06T long in the healthy and M1 cases, while slightly shorter (0.04T) in the model M2. Maps of figure 10 also show that the jet is mainly deflected towards the sinus side (left of the map) in the healthy aorta and towards the commissure side (right of the map) in the most dilated model (M2). In the intermediated case (M1), the jet oscillates nearly symmetrically from the systolic peak, and then it bends leftwards.
The different behaviour of the vorticity released from the valve leaflets is highlighted in figure 11, where the vorticity averaged over the systolic phase is plotted. In the healthy aorta (figure 11a), a red streak slightly inclined rightwards individuates the path followed by the positive vorticity released on the commissure side and advected downstream. Conversely, the negative vorticity released on the opposite side is confined in the sinus. In the most dilated aorta (M2, figure 11c), both positive and negative vorticity released from the leaflets remains within the aortic bulb, with no evidence of meaningful shedding on the left side. An intermediate condition is observed in the case M1, where some positive vorticity is hosted in the bulb, but the signature of significant vortex downstream advection is apparent (figure 11b).

The dynamics of the valve has been measured from the visual inspection of the high-speed images used for velocity measurements. The leaflets begin opening and end closing nearly at the same time irrespective of the dilation (opening starts at $t/T = 0.044 \pm 0.002$ and closing ends at $t/T = 0.432 \pm 0.003$). However, leaflets begin closure earlier in the dilated aortas ($t/T = 0.371, 0.366$ and $0.323$ for model H, M1 and M2, respectively). As a consequence, the closure time is about twice longer in the most dilated model (the duration of the closure is $t/T = 0.063 \pm 0.001$ for model H and M1, and $t/T = 0.111$ for model M2).

## Discussion

The analysis of flow patterns in both the MS and the healthy models shows that the jet originating from the aortic valve exhibits a phase of oscillations during the accelerated ejection, irrespective of the model aorta. Similar oscillations have been described for steady confined jets (Kolšek et al., 2007; Maurel et al., 1996). The non-dimensional period of these oscillations, *i.e.* the Strouhal number, $St = D/(U \cdot T_o)$ ($T_o$ indicates the period of the jet oscillations) is about $St = 0.9$ and does not seem to be meaningfully affected by the dilation of the aortic root.

However, root dilation typical of MS causes a different reorientation of the jet: during the accelerated ejection the jet is mainly orientated towards the side of the sinus in the healthy aorta, and towards the commissure side in the most dilated aorta (M2); whereas oscillations are equally distributed on both sides in the case M1.

The redirection is triggered by the evolution of the vorticity shed from the valve leaflets. On the side of the sinus of Valsalva (right-side of the colour maps, figures 7-9 and 11) the vorticity accumulates in the sinus forming an almost steady vortex, whose size is increased in dilated aortic roots. This behaviour is similar in all three test cases. A different behaviour is observed on the side of the commissure. In the healthy root (H), the vorticity released on the commissure side do not find an expansion in which to form a steady vortex; thus small vortices are advected downstream, partially obstructing the aortic lumen and deflecting the main jet which impinges the vessel wall on the side of the sinus (figures 7a, 11a). Conversely, in the dilated root model M2, the vorticity finds room to accumulate in a steady vortex also on the side of the commissure because in this case the dilation involves the whole root circumference. At the same time, the dilated sinus hosts a larger and stable vortex which tends to deflect the main jet on the opposite (commissure) side (figures 9a, 10a). An intermediate condition is observed in the model M1, where the jet oscillates almost symmetrically on both sides (figures 8a, 8b and 10b). The redirection of the transvalvular jet is consistent with the MRI *in vivo* measurements by Geiger et al. (2013), and possibly triggers the onset of the helical flow in AAo they observe to be correlated with the dilation of the sinuses of Valsalva.

Present experiments were not designed for quantitatively computing wall shear stresses. However, previous computational fluid-dynamics studies, combined with histological analyses, gave evidence of aortic wall thinning and damaging in regions of hemodynamic jet impact, where highest shear stresses are likely to be found (Torii et al., 2013). Therefore, above results indicate that the zone at higher risk of damage progressively moves from the sinus-side, in the healthy aorta, to the commissure-side as the dilation increases. This argument is also supported by the asymmetry in the distributions of shear stresses in the proximal AAo of a healthy volunteer and a MS patient after a valve-sparing aortic replacement evaluated from MRI data (Hope et al., 2013).

The amplitude of the jet oscillations decreases rapidly during the decelerated ejection, and the jet tends to permanently flow on the side of the sinus, which is different in the healthy aorta (sinus side, figure 10a) and in the dilated aortas M1 and M2 (commissure side, figures 10a and 10b).

Comparison of the flow patterns with the dynamics of the valve confirms that the vortices in the dilated sinuses promote an early and smooth closure of the leaflets (Kvitting et al., 2004). In fact, the large, steady supravalvular vortices observed in the case M2 anticipate the beginning of the closure and, correspondingly, nearly double the time taken by the leaflets to close compared to the other cases. The dilation of the aortic annulus, which has been proved to enhance the leaflet stresses, thus increasing the risk of valve injury and malfunction (Grande et al., 2000), is not reproduced in the present experiments.

The altered systolic flow also leads to different residual flow patterns during the diastole. At the end of the systole, the healthy root exhibits an overall counter-clockwise circulation (figure 7c), whereas the circulation is clockwise in both the dilated roots (figures 8c, 9c). It is worth noticing that the variations in the flow pattern may potentially affect the feeding of the coronaries, which occurs during this phase of the cycle.

## Acknowledgements

This work was partially supported by MIUR, grant n. PRIN - 2012HMR7CF. Authors wish to thank Dr. Susanna Grego and Dr. Antonella D'Annolfo for the proficuous discussions and useful suggestions during the design of the experiments.

## Conflicts of interest statement

None

## References

Baumgartner, D., Baumgartner, C., Mátyás, G., Steinmann, B., Löffler-Ragg, J., Schermer, E., Schweigmann, U., Baldissera, I., Frischhut, B., Hess, J., Hammerer, I., 2005. Diagnostic power of aortic elastic properties in young patients with Marfan syndrome. J. Thorac. Cardiovasc. Surg. 129, 730–739. doi:10.1016/j.jtcvs.2004.07.019

Baumgartner, D., Baumgartner, C., Schermer, E., Engl, G., Schweigmann, U., Mátyás, G., Steinmann, B., Stein, J.I., 2006. Different patterns of aortic wall elasticity in patients with Marfan syndrome: A noninvasive follow-up study. J. Thorac. Cardiovasc. Surg. 132, 811–819. doi:10.1016/j.jtcvs.2006.07.001

Brücker, C., Steinseifer, U., Schröder, W., Reul, H., 2002. Unsteady flow through a new mechanical heart valve prosthesis analysed by digital particle image velocimetry. Meas. Sci. Technol. 13, 1043. doi:10.1088/0957-0233/13/7/311

Cafiero, G., Ceglia, G., Discetti, S., Ianiro, A., Astarita, T., Cardone, G., 2014. On the three-dimensional precessing jet flow past a sudden expansion. Exp. Fluids 55, 1–13. doi:10.1007/s00348-014-1677-9

Cecchi, E., Giglioli, C., Valente, S., Lazzeri, C., Gensini, G.F., Abbate, R., Mannini, L., 2011. Role of hemodynamic shear stress in cardiovascular disease. Atherosclerosis 214, 249–256. doi:10.1016/j.atherosclerosis.2010.09.008

Cenedese, A., Prete, Z.D., Miozzi, M., Querzoli, G., 2005. A laboratory investigation of the flow in the left ventricle of a human heart with prosthetic, tilting-disk valves. Exp. Fluids 39, 322–335. doi:10.1007/s00348-005-1006-4

Chandran, K.B., Khalighi, B., Chen, C.-J., 1985. Experimental study of physiological pulsatile flow past valve prostheses in a model of human aorta—II. Tilting disc valves and the effect of orientation. J. Biomech. 18, 773–780. doi:10.1016/0021-9290(85)90052-1

De Tullio, M.D., Cristallo, A., Balaras, E., Verzicco, R., 2009. Direct numerical simulation of the pulsatile flow through an aortic bileaflet mechanical heart valve. J. Fluid Mech. 622, 259–290. doi:10.1017/S0022112008005156

Domenichini, F., Querzoli, G., Cenedese, A., Pedrizzetti, G., 2007. Combined experimental and numerical analysis of the flow structure into the left ventricle. J. Biomech. 40, 1988–1994. doi:10.1016/j.jbiomech.2006.09.024

Espa, S., Badas, M.G., Fortini, S., Querzoli, G., Cenedese, A., 2012. A Lagrangian investigation of the flow inside the left ventricle. Eur. J. Mech. - BFluids 35, 9–19. doi:10.1016/j.euromechflu.2012.01.015

Fortini, S., Querzoli, G., Espa, S., Cenedese, A., 2013. Three-dimensional structure of the flow inside the left ventricle of the human heart. Exp. Fluids 54, 1–9. doi:10.1007/s00348-013-1609-0

Geiger, J., Arnold, R., Herzer, L., Hirtler, D., Stankovic, Z., Russe, M., Langer, M., Markl, M., 2013. Aortic wall shear stress in Marfan syndrome. Magn. Reson. Med. 70, 1137–1144. doi:10.1002/mrm.24562

Geiger, J., Markl, M., Herzer, L., Hirtler, D., Loeffelbein, F., Stiller, B., Langer, M., Arnold, R., 2012. Aortic flow patterns in patients with Marfan syndrome assessed by flow-sensitive four-dimensional MRI. J. Magn. Reson. Imaging 35, 594–600. doi:10.1002/jmri.23500

Grande, K.J., Cochran, R.P., Reinhall, P.G., Kunzelman, K.S., 2000. Mechanisms of aortic valve incompetence: finite element modeling of aortic root dilatation. Ann. Thorac. Surg. 69, 1851–1857. doi:10.1016/S0003-4975(00)01307-2

Gülan, U., Lüthi, B., Holzner, M., Liberzon, A., Tsinober, A., Kinzelbach, W., 2012. Experimental study of aortic flow in the ascending aorta via Particle Tracking Velocimetry. Exp. Fluids 53, 1469–1485. doi:10.1007/s00348-012-1371-8

Hirata, K., Triposkiadis, F., Sparks, E., Bowen, J., Wooley, C.F., Boudoulas, H., 1991. The marfan syndrome: Abnormal aortic elastic properties. J. Am. Coll. Cardiol. 18, 57–63. doi:10.1016/S0735-1097(10)80218-9

Hope, T.A., Kvitting, J.-P.E., Hope, M.D., Miller, D.C., Markl, M., Herfkens, R.J., 2013. Evaluation of Marfan patients status post valve-sparing aortic root replacement with 4D flow. Magn. Reson. Imaging 31, 1479–1484. doi:10.1016/j.mri.2013.04.003

Hutchison, C., Sullivan, P., Ethier, C.R., 2011. Measurements of steady flow through a bileaflet mechanical heart valve using stereoscopic PIV. Med. Biol. Eng. Comput. 49, 325–335. doi:10.1007/s11517-010-0705-z

Kvitting, J.-P.E., Ebbers, T., Wigström, L., Engvall, J., Olin, C.L., Bolger, A.F., 2004. Flow patterns in the aortic root and the aorta studied with time-resolved, 3-dimensional, phase-contrast magnetic resonance imaging: implications for aortic valve–sparing surgery. J. Thorac. Cardiovasc. Surg. 127, 1602–1607. doi:10.1016/j.jtcvs.2003.10.042

Januzzi, J.L., Marayati, F., Mehta, R.H., Cooper, J.V., O'Gara, P.T., Sechtem, U., Bossone, E., Evangelista, A., Oh, J.K., Nienaber, C.A., Eagle, K.A., Isselbacher, E.M., 2004. Comparison of aortic dissection in patients with and without Marfan's syndrome (results from the International Registry of Aortic Dissection). Am. J. Cardiol. 94, 400–402. doi:10.1016/j.amjcard.2004.04.049

Jeremy, R.W., Huang, H., Hwa, J., McCarron, H., Hughes, C.F., Richards, J.G., 1994. Relation between age, arterial distensibility, and aortic dilatation in the Marfan syndrome. Am. J. Cardiol. 74, 369–373. doi:10.1016/0002-9149(94)90405-7

Kolšek, T., Jelić, N., Duhovnik, J., 2007. Numerical study of flow asymmetry and self-sustained jet oscillations in geometrically symmetric cavities. Appl. Math. Model. 31, 2355–2373. doi:10.1016/j.apm.2006.10.010

Leo, H.L., Lim, W.Q.M., Xiong, F.L., Yeo, J.H., 2012. Hemodynamic Investigation of a Stentless Molded Pericardial Aortic Valve, in: Jobbágy, Á. (Ed.), 5th European Conference of the International Federation for Medical and Biological Engineering, IFMBE Proceedings. Springer Berlin Heidelberg, pp. 888–893.

Leopaldi, A.M., Vismara, R., Lemma, M., Valerio, L., Cervo, M., Mangini, A., Contino, M., Redaelli, A., Antona, C., Fiore, G.B., 2012. In vitro hemodynamics and valve imaging in passive beating hearts. J. Biomech. 45, 1133–1139. doi:10.1016/j.jbiomech.2012.02.007

Lim, W.L., Chew, Y.T., Chew, T.C., Low, H.T., 2001. Pulsatile flow studies of a porcine bioprosthetic aortic valve in vitro: PIV measurements and shear-induced blood damage. J. Biomech. 34, 1417–1427. doi:10.1016/S0021-9290(01)00132-4

Loeys, B.L., Dietz, H.C., Braverman, A.C., Callewaert, B.L., De Backer, J., Devereux, R.B., Hilhorst-Hofstee, Y., Jondeau, G., Faivre, L., Milewicz, D.M., Pyeritz, R.E., Sponseller, P.D., Wordsworth, P., De Paepe, A.M., 2010. The revised Ghent nosology for the Marfan syndrome. J. Med. Genet. 47, 476–485. doi:10.1136/jmg.2009.072785

Marom, G., Halevi, R., Haj-Ali, R., Rosenfeld, M., Schäfers, H.-J., Raanani, E., 2013. Numerical model of the aortic root and valve: Optimization of graft size and sinotubular junction to annulus ratio. J. Thorac. Cardiovasc. Surg. 146, 1227–1231. doi:10.1016/j.jtcvs.2013.01.030

Maurel, A., Ern, P., Zielinska, B.J.A., Wesfreid, J.E., 1996. Experimental study of self-sustained oscillations in a confined jet. Phys. Rev. E 54, 3643–3651. doi:10.1103/PhysRevE.54.3643

Meijboom, L.J., Timmermans, J., Zwinderman, A.H., Engelfriet, P.M., Mulder, B.J.M., 2005. Aortic Root Growth in Men and Women With the Marfan's Syndrome. Am. J. Cardiol. 96, 1441–1444. doi:10.1016/j.amjcard.2005.06.094

Mimoun, L., Detaint, D., Hamroun, D., Arnoult, F., Delorme, G., Gautier, M., Milleron, O., Meuleman, C., Raoux, F., Boileau, C., Vahanian, A., Jondeau, G., 2011. Dissection in Marfan syndrome: the importance of the descending aorta. Eur. Heart J. 32, 443–449. doi:10.1093/eurheartj/ehq434

Murdoch, J.L., Walker, B.A., Halpern, B.L., Kuzma, J.W., McKusick, V.A., 1972. Life Expectancy and Causes of Death in the Marfan Syndrome. N. Engl. J. Med. 286, 804–808. doi:10.1056/NEJM197204132861502

Nollen, G.J., Groenink, M., Tijssen, J.G.P., Wall, E.E. van der, Mulder, B.J.M., 2004. Aortic stiffness and diameter predict progressive aortic dilatation in patients with Marfan syndrome. Eur. Heart J. 25, 1146–1152. doi:10.1016/j.ehj.2004.04.033

Pries, A.R., Reglin, B., Secomb, T.W., 2005. Remodeling of Blood Vessels Responses of Diameter and Wall Thickness to Hemodynamic and Metabolic Stimuli. Hypertension 46, 725–731. doi:10.1161/01.HYP.0000184428.16429.be

Pyeritz, R.E., 2000. The Marfan Syndrome. Annu. Rev. Med. 51, 481–510. doi:10.1146/annurev.med.51.1.481

Querzoli, G., Fortini, S., Cenedese, A., 2010. Effect of the prosthetic mitral valve on vortex dynamics and turbulence of the left ventricular flow. Phys. Fluids 1994-Present 22, 041901. doi:10.1063/1.3371720

Reijer, P.M. den, Sallee, D., Velden, P. van der, Zaaijer, E.R., Parks, W.J., Ramamurthy, S., Robbie, T.Q., Donati, G., Lamphier, C., Beekman, R.P., Brummer, M.E., 2010. Hemodynamic predictors of aortic dilatation in bicuspid aortic valve by velocity-encoded cardiovascular magnetic resonance. J. Cardiovasc. Magn. Reson. 12, 4. doi:10.1186/1532-429X-12-4

Richards, A.L., Cook, R.C., Bolotin, G., Buckner, G.D., 2009. A Dynamic Heart System to Facilitate the Development of Mitral Valve Repair Techniques. Ann. Biomed. Eng. 37, 651–660. doi:10.1007/s10439-009-9653-x

Savolainen, A., Keto, P., Hekali, P., Nisula, L., Kaitila, I., Viitasalo, M., Poutanen, V.-P., Standertskjöld-Nordenstam, C.-G., Kupari, M., 1992. Aortic distensibility in children with the Marfan syndrome. Am. J. Cardiol. 70, 691–693. doi:10.1016/0002-9149(92)90215-K

Sturla, F., Votta, E., Stevanella, M., Conti, C.A., Redaelli, A., 2013. Impact of modeling fluid–structure interaction in the computational analysis of aortic root biomechanics. Med. Eng. Phys. 35, 1721–1730. doi:10.1016/j.medengphy.2013.07.015

Teixido-Tura, G., Redheuil, A., Rodríguez-Palomares, J., Gutiérrez, L., Sánchez, V., Forteza, A., Lima, J.A.C., García-Dorado, D., Evangelista, A., 2014. Aortic biomechanics by magnetic resonance: Early markers of aortic disease in Marfan syndrome regardless of aortic dilatation? Int. J. Cardiol. 171, 56–61. doi:10.1016/j.ijcard.2013.11.044

Torii, R., Kalantzi, M., Theodoropoulos, S., Sarathchandra, P., Xu, X.Y., Yacoub, M.H., 2013. Predicting Impending Rupture of the Ascending Aorta With Bicuspid Aortic Valve: Spatiotemporal Flow and Wall Shear Stress. JACC Cardiovasc. Imaging 6, 1017–1019. doi:10.1016/j.jcmg.2013.02.012

Tullio, M.D. de, Pascazio, G., Weltert, L., Paulis, R.D., Verzicco, R., 2011. Evaluation of prosthetic-valved devices by means of numerical simulations. Philos. Trans. R. Soc. Math. Phys. Eng. Sci. 369, 2502–2509. doi:10.1098/rsta.2010.0365

Vismara, R., Leopaldi, A.M., Mangini, A., Romagnoni, C., Contino, M., Antona, C., Fiore, G.B., 2014. In vitro study of the aortic interleaflet triangle reshaping. J. Biomech. 47, 329–333. doi:10.1016/j.jbiomech.2013.11.036

Vitarelli, A., Conde, Y., Cimino, E., D'Angeli, I., D'Orazio, S., Stellato, S., Padella, V., Caranci, F., 2006. Aortic Wall Mechanics in the Marfan Syndrome Assessed by Transesophageal Tissue Doppler Echocardiography. Am. J. Cardiol. 97, 571–577. doi:10.1016/j.amjcard.2005.09.089

Westaby, S., 1999. Aortic dissection in Marfan's syndrome. Ann. Thorac. Surg. 67, 1861–1863. doi:10.1016/S0003-4975(99)00430-0

Yoganathan, A.P., Corcoran, W.H., Harrison, E.C., 1979. In vitro velocity measurements in the vicinity of aortic prostheses. J. Biomech. 12, 135–152. doi:10.1016/0021-9290(79)90153-2

Yun, B.M., Dasi, L.P., Aidun, C.K., Yoganathan, A.P., 2014. Computational modelling of flow through prosthetic heart valves using the entropic lattice-Boltzmann method. J. Fluid Mech. 743, 170–201. doi:10.1017/jfm.2014.54